\documentclass[fleqn,usenatbib,useAMS]{mnras}

\usepackage[utf8]{inputenc} % allow utf-8 input
\usepackage[T1]{fontenc}    % use 8-bit T1 fonts
\usepackage{hyperref}       % hyperlinks
\usepackage{url}    % simple URL typesetting
\usepackage{booktabs}       % professional-quality tables
\usepackage{amsfonts}       % blackboard math symbols
\usepackage{nicefrac}       % compact symbols for 1/2, etc.
\usepackage{microtype}      % microtypography
\usepackage{lineno}
\usepackage{natbib}
\usepackage{pifont}
\usepackage{fleqn}
\usepackage{amsmath,amssymb}
\usepackage{graphicx}
\usepackage{multicol,graphicx,xcolor}
\usepackage{tikz}
\usepackage{verbatim}
\usepackage{bm}
\usetikzlibrary{shapes.misc, positioning}
\usetikzlibrary{matrix, calc, positioning, arrows,shapes,backgrounds, arrows.meta, quotes}
\usepackage{gensymb,stmaryrd,mathtools,textcomp,xspace,grffile}
\usetikzlibrary{shapes.geometric,fit,matrix,positioning,shapes.multipart}
\usetikzlibrary{decorations.markings}
\usetikzlibrary{shadows,decorations.pathreplacing,fadings}
\pgfdeclarelayer{background}
\pgfsetlayers{background,main}
\usetikzlibrary{topaths, calc,3d}
\usetikzlibrary{fit}
\usepackage{color}

\usepackage{bm}		% Bold maths symbols, including upright Greek
\usepackage{pdflscape}	% Landscape pages
\usepackage{ae,aecompl}
\usepackage[section]{placeins}
\usepackage{float}
\usepackage[normalem]{ulem}
\usepackage[toc,page]{appendix}
\usepackage{longtable}

\usepackage[normalem]{ulem}

\usepackage{tabularx}
    \newcolumntype{L}{>{\raggedright\arraybackslash}X}

\usepackage{caption}
\usepackage{subcaption}
\definecolor{bncolor}{RGB}{0,50,255}

\def \apj{Astrophysical Journal}

\definecolor{ndhcolor}{RGB}{204,102,0}

\def \uchicago {University of Chicago, Chicago, IL 60637, USA}
\def \kicp {Kavli Institute for Cosmological Physics, University of Chicago, Chicago, IL 60637, USA}
\def \fnal {Fermi National Accelerator Laboratory, P.O. Box 500, Batavia, IL 60510, USA}
\def \grappa{GRAPPA, University of Amsterdam, Science Park 904,1098 XH Amsterdam, The Netherlands}

\graphicspath{{figures/}}
\begin{document}

\title[KilonovaNet]{\texttt{KilonovaNet}: Surrogate Models of Kilonova Spectra with Conditional Variational Autoencoders}

% format for mnras
\author[Kamilė Lukošiūtė et al.]{
K.~Lukošiūtė$^{1}$,\thanks{Contact e-mail: \href{mailto:k.lukosiute@uva.nl}{k.lukosiute@uva.nl}},
G.~Raaijmakers$^{1}$,
Z.~Doctor$^{2}$,
M.~Soares-Santos$^{3}$,
B.~Nord$^{4,5,6}$
\\
% List of institutions (look at macros.tex
${^1}$\grappa\\
${^2}$Center for Interdisciplinary Exploration and Research in Astrophysics (CIERA),\\Northwestern University, 1800 Sherman Ave, Evanston, IL 60201, USA\\
${^3}$Department of Physics, University of Michigan, Ann Arbor, MI 48109, USA\\
${^4}$\uchicago\\
${^5}$\fnal\\
${^6}$\kicp\\
}
% 
% \author[0000-0001-6082-8529]{M.~Soares-Santos}
% \affil{}

\date{Accepted XXX. Received YYY; in original form ZZZ}
\pubyear{2022}

\maketitle

\begin{abstract}
Detailed radiative transfer simulations of kilonova spectra play an essential role in multi-messenger astrophysics. 
Using the simulation results in parameter inference studies requires building a surrogate model from the simulation outputs to use in algorithms requiring sampling. 
In this work, we present \texttt{KilonovaNet}, an implementation of conditional variational autoencoders (cVAEs) for the construction of surrogate models of kilonova spectra.
This method can be trained on spectra directly, removing overhead time of pre-processing spectra, and greatly speeds up parameter inference time.
We build surrogate models of three state-of-the-art kilonova simulation data sets and present in-depth surrogate error evaluation methods, which can in general be applied to any surrogate construction method. 
By creating synthetic photometric observations from the spectral surrogate, we perform parameter inference for the observed light curve data of GW170817 and compare the results with previous analyses.
Given the speed with which \texttt{KilonovaNet} performs during parameter inference, it will serve as a useful tool in future gravitational wave observing runs to quickly analyze potential kilonova candidates.

\end{abstract}

\section{Introduction}

%%%%%%%%%%%%%%%%%%%%%%%%%%%%%%%%%%%%%%%%
%%% Set the stage and scope of the introduction:
%%% what can we learn from merger events and how can we learn it?
%%%%%%%%%%%%%%%%%%%%%%%%%%%%%%%%%%%%%%%%
% GW's EMs, mergers, what are they generally useful for?
Mergers of compact stellar remnants, like neutron stars (NSs) and black holes (BHs), provide important testbeds for astrophysical processes, cosmological evolution, and matter under extreme conditions.
In binary merger systems, as companions orbit each other, their physical separation decays due to energy loss through the emission of gravitational waves (GWs), which can be detected by modern ground-based interferometers, like those of the Laser Interferometer Gravitational-Wave Observatory \citep[LIGO;][]{AdvancedLIGO}, Virgo \citep{AdvancedVirgo}, and Kamioka Gravitational Wave Detector \citep[KAGRA;][]{KAGRA} experiments.
If the merging system contains one or two neutron stars, neutron star material can be ejected at high velocities and emit electromagnetic (EM) waves \citep{lattimerBlackholeneutronstarCollisions1974,liTransientEventsNeutron1998,MetzgerBergerPromising, 10.3389/fspas.2020.609460}.
Using information jointly from GWs and EM emission is essential for maximizing our understanding of multiple phenomena: since both GW and EM measurements are sensitive to parameters of a merging binary, combining them can yield stronger constraints on properties of the binary \citep{CoughlinTowardRapidKN}, the neutron-star equation of state \citep{Radice:2017lry,coughlinConstraintsNeutronStar2018}, and the expansion rate of the universe \citep{Schutz1986,Holz:2005df,abbottGravitationalwaveStandardSiren2017}. 

\begin{figure*}
    \centering
    \includegraphics[width=0.9\textwidth]{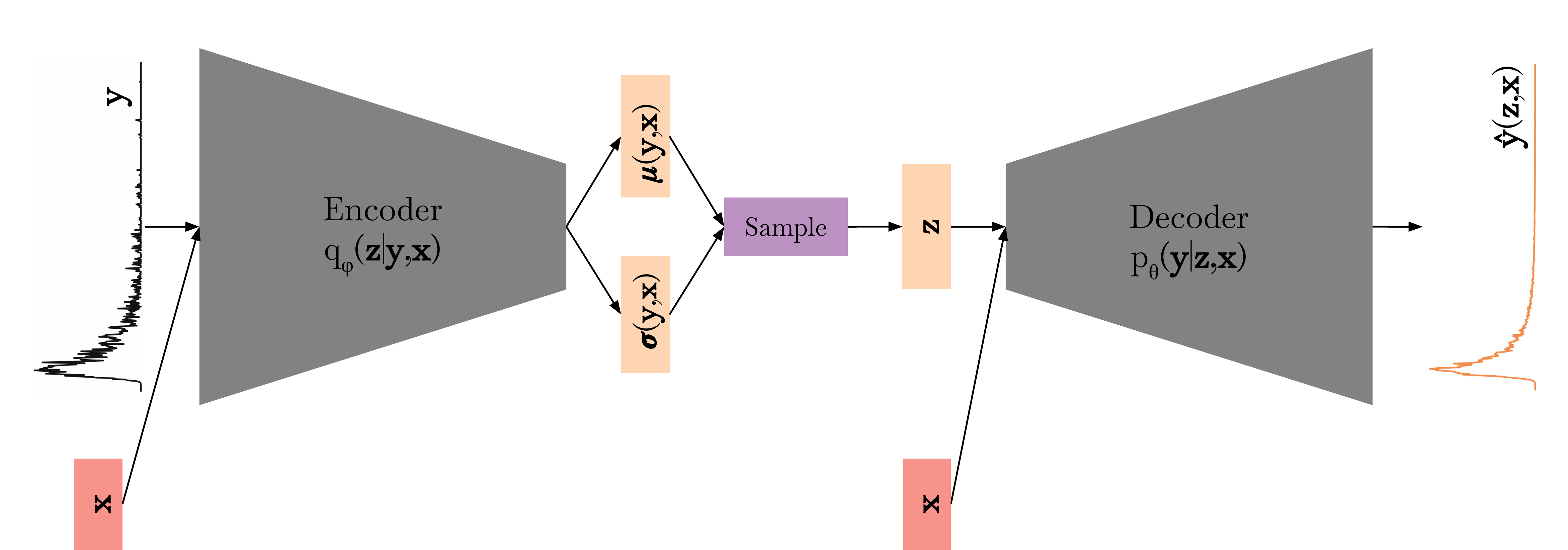}
    \caption{
    % describe each component of VAE - encoder, latent, decoder, ...
    % 
    Diagram of the cVAE architecture. There are two neural networks -- encoder (left, gray) and  decoder (right, gray -- and a latent space ($\mathbf{z}$, pink). 
    The inputs to the encoder are the physical parameter  $\mathbf{x}$ and the spectrum $\mathbf{y}$ at time step $t$ for a single kilonova event. 
    % pairs of spectra $\mathbf{y}$ and corresponding physical parameters $\mathbf{x}$.
    The outputs from the encoder are  $\mathbf{\sigma}(\mathbf{x, y})$ and $\boldsymbol{\mu}(\mathbf{x, y})$, which are the parameters of the distribution $q_{\phi}(\mathbf{z}|\mathbf{y}, \mathbf{x})$ over the latent variables $\mathbf{z}$.
    % , which are themselves passed to the decoder along with the input physical parameters $\mathbf{x}$.
    The inputs to the decoder are $\mathbf{z}$ and $\mathbf{x}$, which are used to draw a sample from the distribution $p_{\theta}(\mathbf{y} | \mathbf{z},\mathbf{x})$ over the spectra. This produces an output $\hat{y}(z,x)$, a new surrogate spectrum. %and the output is are the .
     After training, the decoder is used as the generator of simulated data for the surrogate model: it takes as input a sample from $\mathcal{N}(0, \mathbf{I})$ and a set of physical parameters $\mathbf{x}$ and predicts a spectrum $\hat{\mathbf{y}}(\mathbf{z}, \mathbf{x})$.
    }
    \label{fig:cvae}
    
\end{figure*}

%%%%%%%%%%%%%%%%%%%%%%%%%%%%%%%%%%%%%%%%
%%% what are merger events and their associated novae?
%%%%%%%%%%%%%%%%%%%%%%%%%%%%%%%%%%%%%%%%
Amidst the neutron-rich material ejected during a NS-containing merger \citep{lattimerBlackholeneutronstarCollisions1974,rosswogMassEjectionNeutron1998,metzgerElectromagneticCounterpartsCompact2010}, r-process nucleosynthesis generates heavy nuclei \citep{burbidgeSynthesisElementsStars1957,cameronNUCLEARREACTIONSSTARS1957}.
The subsequent decay of these nuclei drives a \emph{kilonova} --- an ultraviolet-optical-infrared transient, whose brightness peaks two to three days after a merger \citep{liTransientEventsNeutron1998}.
%%% types of mergers
Most binary neutron star (BNS) systems and some black hole - neutron star (BHNS) systems are expected to produce a kilonova \citep{MetzgerKNHandbook}.

There are two primary channels for mass-ejection during a NS-containing merger \citep{MetzgerKNHandbook} --- dynamical ejection (during merger) and disk wind ejection (after merger).
Dynamical ejecta are further subdivided into two categories -- ``tidal'' and ``polar.''
Tidal dynamical ejecta outflows occur during the final stage of the inspiral in both BHNS and BNS systems when the neutron star is disrupted by the gravitational field of the companion.
Polar dynamical ejecta result from shock heating caused by the direct collision of the neutron stars in BNS systems only \citep[e.g.,][]{radiceBinaryNeutronStar2018}.

%%%%%%%%%%%%%%%%%%%%%%%%%%%%%%%%%%%%%%%%
%%% what is the history of observing these events?
% describe kilonovae
%%% detecting GWs and kilonovae
%%%%%%%%%%%%%%%%%%%%%%%%%%%%%%%%%%%%%%%%
The detection of GW170817, the first observed BNS GW event, marks the onset of the era of multi-messenger and GW astronomy \citep{ligoscientificcollaborationandvirgocollaborationGW170817ObservationGravitational2017}.
GW170817 was accompanied by counterparts across the EM spectrum \citep[e.g.,][]{arcaviOpticalEmissionKilonova2017,coulterSwopeSupernovaSurvey2017,droutLightCurvesNeutron2017,shappeeEarlySpectraGravitational2017,smarttKilonovaElectromagneticCounterpart2017,soares-santosElectromagneticCounterpartBinary2017,valentiDiscoveryElectromagneticCounterpart2017,abbottMultimessengerObservationsBinary2017, goldsteinGROWTHS190426cRealtime2019, savchenkoINTEGRALDetectionFirst2017,abbottGravitationalWavesGammaRays2017}.
Then during the third LIGO and Virgo observing run (Run O3), tens of binary black hole mergers and multiple neutron star-containing mergers were detected, including BHNS events \citep{abbottGWTC2CompactBinary2021,NSBHLIGO,GWTC3}. 
% potentially containing neutron stars were detected
%Additionally, during the second half of the third run (Run O3b), two events consistent with BHNS mergers were detected \citep{NSBHLIGO}.  
Through multiple dedicated follow-up electromagnetic observing campaigns in O3 \citep{coughlinGROWTHS190425zSearching2019, goldsteinGROWTHS190426cRealtime2019, antierGRANDMAObservationsAdvanced2020,pageSwiftXRTFollowupGravitational2020,gompertzSearchingElectromagneticCounterparts2020,Anand:2020eyg}, 
one candidate was identified  \citep{grahamCandidateElectromagneticCounterpart2020}, but zero were confirmed.

%%%%%%%%%%%%%%%%%%%%%%%%%%%%%%%%%%%%%%%%
%%% How do we study these signals 
%%%%%%%%%%%%%%%%%%%%%%%%%%%%%%%%%%%%%%%%

Accurate and fast models of kilonovae are critical for multiple tasks in multimessenger and GW astrophysics.
First, we need fast models in order to quickly evaluate candidates during electromagnetic follow up campaigns \citep{soares-santosElectromagneticCounterpartBinary2017}. 
Identifying and observing a candidate in the first few hours is crucial, especially to understand the origin of early kilonova emission, such as the fast, blue emission seen in GW170817 \citep{ArcaviFirstHoursGW170817}.
After the data is all collected, analysis, inference can begin; we can use the data to infer parameters of the NS equation of state \citep[e.g.,][]{radiceBinaryNeutronStar2018, coughlinConstraintsNeutronStar2018, RaaijmakersPSRj0740}, the Hubble constant \citep[e.g.,][]{abbottGravitationalwaveStandardSiren2017, coughlinMeasuringHubbleConstant2020, dietrichMultimessengerConstraintsNeutronstar2020}, and studying r-process nucleosynthesis \citep[e.g.,][]{droutLightCurvesNeutron2017} and the site of rapid neutron capture nucleosynthesis \citep[e.g.,][]{kasenOriginHeavyElements2017}. 
For these scientific studies, the most accurate models will lead to the least biased results.

%%%%%%%%%%%%%%%%%%%%%%%%%%%%%%%%%%%%%%%%
%%% background on simulations and why they're expensive
%%%%%%%%%%%%%%%%%%%%%%%%%%%%%%%%%%%%%%%%
Numerical hydrodynamic simulations provide the most detailed models of matter (i.e., velocity, mass, composition, and opacity) ejected during mergers 
\citep[e.g.,][]{dietrichModelingDynamicalEjecta2017, perego2017gfoAnisotropicThreecomponent2017, kawaguchiLowmassBinaryNeutron2021, tanakaSystematicOpacityCalculations2020}.
The results of these simulations are then incorporated into radiative transfer calculations to compute observational properties of mergers \citep[e.g.,][]{bullaPOSSISPredictingSpectra2019, kasenOriginHeavyElements2017, kawaguchiModelsKilonovaMacronova2016} to generate synthetic observations for comparison with real-sky observations \citep[e.g.,][]{raaijmakersChallengesAheadMultimessenger2021}.
Since the simulations are computationally complex, they take several hours to produce observables for one parameter set \citep{bullaPOSSISPredictingSpectra2019}).

In order to use the outputs of the simulations in parameter inference studies, the outputs are trained to produce a ``surrogate" model that emulates the output of simulations but run considerably faster.
For example, a commonly used surrogate-construction method is Gaussian process regression (GPR), a statistical method for interpolation; this method has been applied in a few GW and EM studies \citep{doctorStatisticalGravitationalWaveform2017, coughlinConstraintsNeutronStar2018, dietrichMultimessengerConstraintsNeutronstar2020}. 
One difficulty with using GPR for kilonova surrogate models is that it does not scale well with the number of training examples: the evaluation time goes as the square of the number of training points and the training time goes as the cube. 
State-of-the-art radiative transfer simulations, such as those published in \citet{dietrichMultimessengerConstraintsNeutronstar2020} and \citet{anandOpticalFollowupNeutron2021}, output spectra based on multiple input parameters in hundreds of frequency bins and with hundreds of time steps. 
The high-dimensionality of the inputs and outputs, combined with the grid over physical parameters means that the simulated sets become large quickly, making GPR a costly method.
Recently, \cite{almuallaUsingNeuralNetworks2021} presented a neural network based surrogate modelling framework which greatly increased inference time for the GW170817 light curve. 
We present a complementary approach to their method, instead training directly on spectra, thereby skipping the dimensional reduction step, and training one permanent network set, thereby allowing construction of photometric observations for any band from the surrogate.

\texttt{KilonovaNet} is an algorithm for generating surrogate kilonova spectra on tens-of-millisecond timescales via conditional variational autoencoders (cVAE) \citep{ kingmaAutoEncodingVariationalBayes2014, rezendeStochasticBackpropagationApproximate2014, sohnLearningStructuredOutput2015}.
We additionally present a thorough error evaluation suite to understand the performance of \texttt{KilonovaNet} across the input parameter space; these error evaluation methods can also be applied to any future surrogate construction methods. The code used to produce the results in our study is available at \url{https://github.com/klukosiute/kilonovanet}.

The work is organized as follows. 
In Section \ref{sec:data}, we discuss the simulated data sets used for constructing the surrogate model.
In Section \ref{sec:method}, we discuss the cVAE, its limitations with respect to uncertainty estimation, and our approach for quantifying uncertainties.   
We discuss the performance of the cVAE surrogate model in Section \ref{sec:results} and conclude in Section \ref{sec:conclusion}.

%%%%%%%%%%%%%%%%%%%%%%%%%%%%%%%%%%%%%%%%%%%%%%%%%%%%%%%%%%%%%%%%
%%%%%%%%%%%%%%%%%%%%%%%%%%%%%%%%%%%%%%%%%%%%%%%%%%%%%%%%%%%%%%%%
\section{Data}
\label{sec:data}
In this work, we report the construction and testing of surrogate models for three publicly available datasets of simulated kilonova -- in particular, spectra of the material ejected during a BNS or BHNS merger:
the \emph{Kasen} BNS simulations~\citep{kasenOriginHeavyElements2017}, the \emph{Dietrich} BNS simulations~\citep{dietrichMultimessengerConstraintsNeutronstar2020}, and the \emph{Anand} BHNS simulations~\citep{anandOpticalFollowupNeutron2021}.
In the \emph{Kasen} simulation, the key physical parameters of the single-component spherically symmetrically expanding ejecta are the ejecta mass $M_{ej}$, the characteristic expansion velocity of the ejecta $v_{ej}$, and the chemical composition as indicated by the lanthanide fraction $\chi$. 
Each spectrum is computed in increments of 0.1 days, from 0.1 to 24.9 days post-merger and has 1629 logarithmically spaced wavelength bins from 149 \AA\ to 99467 \AA.
There are two sets of publicly-available parameter grids for these simulations available: a regularly spaced grid, whose parameter values are given in Table \ref{table:params}, and a narrower, irregularly-spaced set. The values of the second set lie mostly within the ranges of the regularly spaced grid but also includes several models with $v_{ej} / c = 0.4$. 
This set was was produced particularly for the study of GW170817. 
There are two sets of parameters. 
We combine the regular grid data (329 parameter sets) and the narrower, irregular grid data (22 parameter steps) to create our full training dataset, resulting in 351 unique parameter combinations; with 249 time steps per parameter set, this leads to a total of 87399 spectra in the \emph{Kasen} simulated dataset.

The \emph{Dietrich} BNS simulations are generated with \textsc{possis}, a multi-dimensional Monte Carlo radiative transfer code \citep{bullaPOSSISPredictingSpectra2019}. 
The parameter sets consist of the mass of the dynamical ejecta $M_{ej, dyn}$, the mass of the post-merger ejecta $M_{ej, pm}$, the half-opening angle of the lanthanide-rich tidal dynamical ejecta $\Phi$, and the cosine of the observer viewing angle $\cos \theta_{obs}$.
The half-opening angle indicates a separation of the ejecta into three components: the matter above and below the ejecta are lanthanide-free.
Each spectrum is computed at increments of 0.2 days, starting at 0.2 days up to 20 days post-merger and for 500 evenly spaced wavelength bins from 100 \AA\ to 99,900 \AA. 
There are in total 2156 combinations of the parameter values listed in Table \ref{table:params}.
With 100 time steps for each parameter combination, this leads to 215,600 spectra for this simulation set. 

The \emph{Anand} BHNS data set is also generated through \textsc{possis}.
The dynamical ejecta are concentrated within an angle $\phi = 30 \deg$ above and below the equatorial plane, expanding with velocities ranging 0.1$c$ to 0.3$c$, and with a lanthanide-rich composition. 
The post-merger ejecta are assumed to be spherical, expanding with velocities ranging 0.025$c$ to 0.1$c$, and with an intermediate lanthanide composition. 
Table \ref{table:params} lists all the values of these parameters as well as the final parameter $\cos \theta_{obs}$. 
There are 891 unique parameter combinations and again with 100 time steps each.

\begin{table}
\caption{Parameter values for each of the three kilonova simulation data sets which we use for training our surrogate model.  }
\begin{tabularx}{\linewidth}{LL} 

\hline \hline
\textbf{Kasen et al. 2017}         &                                                             \\ \hline 
Parameters                & Values                                                     \\ \hline
$M_{ej} / M_{\odot}$      & $\{ 0.001, 0.0025, 0.005,$ $0.0075, 0.01, 0.25, 0.05, 0.1 \}$ \\
$v_{ej} / c $             & $\{ 0.03, 0.05, 0.1, 0.2, 0.3\}$                            \\
$\log_{10} \chi $         & $ \{ -9, -5, -4, -3, -2, -1 \}$                             \\ \hline

\textbf{Dietrich et al. 2020 }     &                                                             \\ \hline 

$M_{ej, dyn} / M_{\odot}$ & $\{0.001, 0.005,  0.01, 0.02\}$                             \\
$M_{ej, pm} / M_{\odot}$  & $\{0.01, 0.03, 0.05, 0.07, $ $0.09, 0.11, 0.13\}$              \\
$\Phi$ (deg)              & $\{ 0, 15, 30, 45, 60, 75, 90\}$                            \\
$\theta_{obs}$            & $\{ 0, 0.1, 0.2, 0.3, 0.4, 0.5,$ $0.6, 0.7, 0.8, 0.9, 1 \}$   \\ \hline
\textbf{Anand et al. 2021}         &                                                             \\ \hline

$M_{ej, dyn} / M_{\odot}$ & $\{0.01, 0.02, 0.03,0.04,0.05,$ $0.06, 0.07,0.08, 0.09\}$       \\
$M_{ej, pm} / M_{\odot}$  & $\{0.01, 0.02, 0.03,0.04 0.05,$ $0.06 0.07,0.08,0.09\} $       \\
$\cos \theta_{obs}$       & $\{ 0, 0.1, 0.2, 0.3, 0.4, 0.5, $ $0.6, 0.7, 0.8, 0.9, 1 \}$   \\
\hline \hline
\end{tabularx}
\label{table:params}
\end{table}

%%%%%%%%%%%%%%%%%%%%%%%%%%%%%%%%%%%%%%%%%%%%%%%%%%%%%%%%%%%%%%%%
%%%%%%%%%%%%%%%%%%%%%%%%%%%%%%%%%%%%%%%%%%%%%%%%%%%%%%%%%%%%%%%%
\section{Methods}
\label{sec:method}

We use the conditional variational autoencoder (cVAE) to construct our surrogate models. 
We then highlight the subtle but critical challenge of deriving uncertainty estimates directly from Gaussian cVAE's, caused by variance shrinkage. 
Finally, we discuss data pre-processing, our training protocol, and our method for evaluating the trained cVAE-based model.

\subsection{Conditional Variational Autoencoder}
We start with a simplified derivation of the variational autoencoder \citep[VAE;][]{kingmaAutoEncodingVariationalBayes2014}, which is a generative model that relies on variational Bayesian methods for optimization.
We note the following definitions:

\begin{itemize}
\item $\boldsymbol{x}$: vector of physical parameters, including those discussed in \S~\ref{sec:data}, like ejecta mass, velocity, and lanthanide fraction, as well as the timestep $t$ in the evolution a single kilonova.
\item $\boldsymbol{y}$: vector of spectral data from a single kilonova.
\item $\boldsymbol{z}$: vector of latent variables
\item $p$: generative model
\item $q$: inference model
\item $\theta$: parameters for generative neural model network\footnote{$\theta_{obs}$ in \S\ref{sec:data} is unrelated to $\theta$.}
\item $\phi$: parameters for inference model neural network
\end{itemize}

Our goal is to obtain a model to efficiently approximate a generative process $p^{*}(\boldsymbol{y})$ (the radiative transfer model) to predict a spectrum $\boldsymbol{y}$.
We seek $p_{\theta}(\boldsymbol{y}) = \int p_{\theta}(\boldsymbol{y}|z) dz$ to approximate $p^{*}(\boldsymbol{y})$, where $p_{\theta}(y|z)$ is part of a deep latent variable model (DLVM) and is related to it via the definition of conditional probability.
To evaluate the integral, an inference model $q_{\phi}(\mathbf{z} | \mathbf{y})$ (the encoder) is used to approximate the true $p_{\theta} (\mathbf{z} | \mathbf{y})$, which, through Bayes' theorem, approximates $p_{\theta}(\mathbf{y} | \mathbf{z})$.
A VAE is a deep neural network-based framework for co-optimizing a DLVM and an inference model.

A model $EncoderNeuralNet$ is trained to return the mean $\boldsymbol{\mu}$ and the variance $\boldsymbol{\sigma}$ of a multidimensional Gaussian for a particular data input $\mathbf{y}$. 
The mean and variance are then used for sampling the generative model: 

\begin{equation}
\begin{aligned}
(\boldsymbol{\mu}, \log \boldsymbol{\sigma}) 
&=\text {EncoderNeuralNet}_{\phi}(\mathbf{y}) \\
q_{\phi}(\mathbf{z} | \mathbf{y}) &=\mathcal{N}(\mathbf{z} ; \boldsymbol{\mu}, \operatorname{diag}(\boldsymbol{\sigma})),
\end{aligned}
\end{equation}
\noindent  

The optimization objective of the VAE is the Evidence Lower Bound (ELBO), which allows the simultaneous optimization of the parameters $\theta$ (generative model), and the variational parameters $\phi$ (inference model).
The loss function is defined as
\begin{equation}
\mathcal{L}_{\theta, \phi}(\mathbf{y}) = - D_{KL}(q_{\phi}(\mathbf{z} | \mathbf{y}) || p_{\theta}(\mathbf{z})) + \mathbb{E}_{q_{\phi}(\mathbf{z}|\mathbf{y})} [ ( p_{\theta}(\mathbf{y | \mathbf{z}}) ) ], 
\end{equation}
\noindent where $D_{KL}$ is the Kullback-Liebler Divergence, which measures the distance between the probability $p$ and the inference model $q$, and $\mathbb{E}$ is the reconstruction loss.
Using Monte Carlo sampling and the ``reparametrization trick,'' the networks $p_{\theta}$ and $q_{\phi}$ can be optimized using gradient descent methods \citep{kingmaAutoEncodingVariationalBayes2014, kingmaIntroductionVariationalAutoencoders2019, rezendeStochasticBackpropagationApproximate2014}. 

The variational autoencoder learns an approximation for the distribution  $p^*(\mathbf{y})$, and the decoder returns the parameters of the distribution $p_{\theta}(\mathbf{y} | \mathbf{z})$.
After training, the marginal likelihood $p(\mathbf{y})$ can be estimated through importance sampling, where random samples are drawn from $q_{\phi}(\mathbf{z}| \mathbf{y})$ \citep{rezendeStochasticBackpropagationApproximate2014}:
\begin{equation}
p_{\boldsymbol{\theta}}(\mathbf{y})= \mathbb{E}_{q_{\phi}(\mathbf{z} \mid \mathbf{y})}\left[p_{\boldsymbol{\theta}}(\mathbf{y}, \mathbf{z}) / q_{\phi}(\mathbf{z}|\mathbf{y})\right] \approx \frac{1}{L} \sum_{l=1}^{L} \frac{ p_{\boldsymbol{\theta}}(\mathbf{y}|\mathbf{z}^{(l)}) p(\mathbf{z}^{(l)})}{q_{\phi}(\mathbf{z}|\mathbf{y})}
\end{equation}
Through this process, the VAE can learn a  multi-dimensional distribution as a data-driven model. 
This distribution $p_{\theta}(\mathbf{y} | \mathbf{z})$ is then the approximation of the $p^*(\mathbf{y})$, the true generative model from the radiative transfer simulation.

In this work, we aim to learn a distribution over the spectral data $\mathbf{y}$. 
But, importantly, we must condition the VAE model on input physical parameters $\mathbf{x}$ of the kilonova, transforming the model into a cVAE~\citep[][]{sohnLearningStructuredOutput2015, kingmaIntroductionVariationalAutoencoders2019}: $p_{\theta}(\mathbf{y}|\mathbf{x})$.
The optimization objective for the cVAE is the conditional likelihood
\begin{equation}
\begin{aligned}
\mathcal{L}_{\theta, \phi}(\mathbf{y} | \mathbf{x}) &= - D_{KL}(q_{\phi}(\mathbf{z} | \mathbf{y}, \mathbf{x}) || p_{\theta}(\mathbf{z} | \mathbf{x})) \\
&+ \mathbb{E}_{q_{\phi}(\mathbf{z}|\mathbf{y}, \mathbf{x})} [ ( \log p_{\theta}(\mathbf{y | \mathbf{z}, \mathbf{x}}) ) ].
\end{aligned}
\label{eq:cvae_objective}
\end{equation}
A graphical representation of the model is shown in Fig.~\ref{fig:cvae}.

Since sampling from $p(\mathbf{z}) = \mathcal{N}(0, \mathbf{I})$ produces negligible differences in the prediction of new spectra $\mathbf{y}$, we choose to always use the  same sample (the central value of $\mathcal{N}(0, \mathbf{I})$, a vector of zeroes) for $\mathbf{z}$ when predicting spectra.

\subsection{Variance Shrinkage of Gaussian VAEs }
Using the prescription above, cVAEs can generate multi-dimensional data distributions conditioned on input variables when an appropriate likelihood function is chosen.
For binary classification, the Bernoulli distribution is typically used to describe data that have binary outcomes.
The log-likelihood is given by the binary cross-entropy (BCE) loss,
\begin{equation}
    \log p_{\theta}(\mathbf{y | \mathbf{z}}) = -\sum_{i=1}^{D}  y_i \log \hat{y}_i + (1-y_i)\log (1-\hat{y}_i),
    \label{eq:bernoulli_nll}
\end{equation}
\noindent where $D$ is the dimension of the vector to be predicted, $y_i$ is a true label, and $\hat{y}_i$ is a predicted label. 

For real-valued data, like the kilonova spectra in this work, the multivariate Gaussian with a diagonal covariance offers flexibility and mathematical simplicity with a log-likelihood of the form 
\begin{equation}
    \log p_{\theta}(\mathbf{y | \mathbf{z}}) = \sum_{i=1}^{D}  -\frac{1}{2\sigma_{\theta, i}^2(z)} || y_i - \mu_{\theta, i}(z) ||^2 - \frac{1}{2} \log 2 \pi \sigma_{\theta, i}^2(z),
    \label{eq:gauss_nll}
\end{equation}
\noindent where $\sigma_{\theta, i}^2(z)$ and  $\mu_{\theta,i}(z)$ are the variance and mean, respectively, for the $i$-th dimension.

If we used the Gaussian likelihood for our continuous and real data outputs, we could consider interpreting the output distribution as an uncertainty learned by the model.
However, it has recently been shown that the maximum likelihood objective is ill-posed for continuous models, such as those employing Gaussian distributions \citep{matteiLeveragingExactLikelihood2018}.
For models trained so that $y_i\sim\mu_{\theta, i}(z)$, the $-\frac{1}{2} \log 2 \pi \sigma_{\theta, i}^2(z)$ term will push the variance to zero before the $\frac{1}{2 \sigma_{\theta, i}^2(z)}$ term can catch up.
This can be seen in Equation~\ref{eq:gauss_nll}. 
Therefore, a Gaussian cVAE will produce a small variance that has doesn't have useful physical interpretation.
This ``variance shrinkage problem'' prohibits use of the cVAE model as a probabilistic method  \citep{matteiLeveragingExactLikelihood2018, detlefsenReliableTrainingEstimation2019}.

There are several standard procedures to avoid the variance shrinkage problem in practice \citep{detlefsenReliableTrainingEstimation2019}.
For example, setting a globally constant variance -- e.g., $\sigma^2 = 1$ -- the log-likelihood becomes the mean squared error, and the cVAE loss function becomes
\begin{equation}
    \log p_{\theta}(\mathbf{y | \mathbf{z}}) = \sum_{i=1}^{D}  -|| y_i - \mu_{\theta, i}(z) ||^2. 
\end{equation}
This uncertainty is predetermined and not learned from the data. 

Because the MSE loss does not provide statistical interpretability, the Bernoulli distribution is often used for non-binary data: optimizing a Bernoulli log-likelihood (the BCE loss in Equation \ref{eq:bernoulli_nll}) is considerably simpler than optimizing the MSE \citep{detlefsenReliableTrainingEstimation2019}.
This can be done when the data is scaled to the range $[0,1]$.
However, the outputs will not have a meaningful statistical interpretation because the Bernoulli distribution is meant for discrete random variables with a binary outcome.
Our initial experiments verify that training with BCE loss outperforms the MSE loss in terms of accuracy and efficiency. 

The closed-form expression for our optimization objective is thus
\begin{multline}
\mathcal{L} = -\frac{1}{2}\sum_{i=1}^{D}\left[ 1 + \log \sigma_i^2  - \sigma_i^2 - \mu_i^2 \right]\\ -\sum_{i=1}^{D}  y_i \log \hat{y}_i+ (1-y_i)\log (1-\hat{y}_i),
\label{eq:exact_objective}
\end{multline}
\noindent where we have used the BCE loss and  $p(\mathbf{z}|\mathbf{x}) = \mathcal{N}(0, I)$, as per the definition of the VAE model, and  where $\mu_i$ and $\sigma^2_i$ are the outputs of the encoder distribution for passing a data point pair $(x_i, y_i)$, with a resulting prediction $\hat{y}_i$.
While variance shrinkage prevents learning of the variance from the data, the cVAE can still reproduce spectral data with high fidelity.
We pursue estimates of the variance post-facto through measurements of the surrogate model data with respect to the original simualted data.

\subsection{Data Pre-processing}
\label{sec:datapreproc}
The original \emph{Kasen} spectra are in units of erg s$^{-1}$ Hz$^{-1}$, and the \emph{Dietrich} and \emph{Anand} spectra are in units of erg s$^{-1}$ cm$^{-2}$ \AA$^{-1}$ at a distance of 10 pc. 
To use the data sets for network training and to evaluate the surrogate results, we first process all spectra such that they have units erg s$^{-1}$ \AA$^{-1}$.
We then re-scale the input physical parameters and the spectra to the range $[0,1]$ so that the data lie in the supported range of the sigmoid activation function in the final layer of the decoder neural network. 
Finally, we separate each of the three simulation data sets into training, validation, and test sets with relative proportions 80:10:10, respectively.
We use only the training and validation sets to optimize the training and hyperparameters.

\subsection{Training}
\label{sec:training}

For each of the three simulated kilonova spectra data sets, we train a distinct cVAE model with the following procedure. 
\begin{enumerate}
    \item We use the training and validation data sets from the top-level split to perform hyperparameter selection for our final architecture. 
    For each simulation data set, we perform a hyperparameter search using one of the train-validation data splits.
    We consider the hyperparameters of the cVAE model to be the dimensionality of the latent space $\mathbf{z}$ and the dimensionality of the hidden layers of the decoder and encoder.
    Because the encoder is considered the approximate inverse of the decoder, we use one hyperparameter for both the hidden layers.
    We perform hyperparameter selection only on these two values; all other hyperparameters of the model, such as the learning rate and batch size, are fixed across all of our tests and models. 
    We trained 16 architectures with four values for each hyperparameter.
    \item We then split the training and validation sets into nine total sets, and train the architecture on these subsets. 
    \item We then evaluate all nine models on the test set, which allows us to develop a statistical analysis of the model's predictive capability.
    \item We chose one of the nine models at random for the final production model that we use to show examples and make publicly available. 
\end{enumerate}

We train each model -- both when performing hyperparameter selection and when training each of the nine data split models -- for 200 epochs. 
The final training losses for the \emph{Kasen}, \emph{Dietrich}, and \emph{Anand} data, averaged over the nine data split models, respectively, are 48.54, 11.23, and 8.20.
The training during the hyperparameter search required $\sim24$ hours, and the final nine experiment models required approximately 26 hours of training on the same GPU -- both on an Nvidia GeForce 1080Ti GPU.
The prediction of 100 spectra of the trained network from unique parameter sets requires approximately 10 milliseconds on one Intel® Core™ i7-7700HQ CPU. 
We use PyTorch for implementation \citep{paszkePyTorchImperativeStyle2019} and Adam as our gradient-based optimizer \citep{kingmaAdamMethodStochastic2017}.

\subsection{Model Evaluation and Quantifying Prediction Uncertainties}
\label{sec:err_analysis}

For each simulation dataset, we have constructed a surrogate cVAE-based model with which we can generate data under the supported ranges of the physical input variables $\mathbf{x}$ and output spectra $\mathbf{y}$.

We next seek to generate uncertainties associated with the predictions of the spectra.
Variance shrinkage and the resulting training choices dictate that the cVAE cannot produce statistically meaningful variances. 
Therefore the errors in the resulting surrogate models still need to be characterized.
We use the cVAE only as a method to produce more samples of the original radiative transfer simulations with which the cVAE was trained: we do not seek to improve upon or extrapolate beyond those simulations.
In the following sections, we discuss error sources and procedures for quantifying them.

\subsubsection{Sources of Error}

We examined and estimated multiple sources of error with respect to the original simulations.
First, for systematic error, we account for the bias of the predicted spectra as a function of input parameters by comparing predictions to the truth generated by the radiative transfer simulations.
The surrogate model is generated to minimize bias in reconstructing spectra, so the model (with its chosen hyperparameter values) is designed to minimize this bias.
Second, we account for statistical error of the predictions over the space of input parameters and different trained models, as we discuss in detail below.

\begin{figure}
    \centering
    \includegraphics[width=0.45\textwidth]{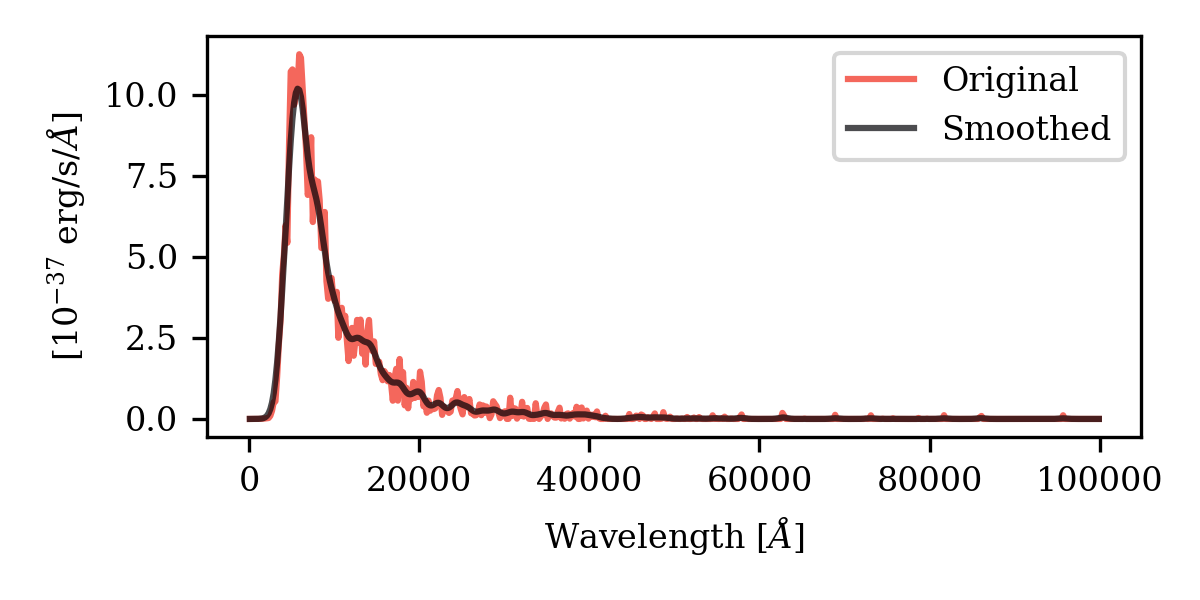}
    \caption{An example original spectrum (orange) and its smoothed version using a Gaussian kernel with $\sigma=2.5$ (blue). This spectrum is from the \emph{Dietrich} data set and corresponds to the parameters $M_{ej, dyn}/M_{\odot} = 0.01$, $M_{ej, pm}/M_{\odot} = 0.07$, $\Phi = 30.0^{\circ}$, $\cos$ $\Theta_{obs} = 0.5$, $t = 1.8$ days.}
    %  that we use to estimate the size of the inherent stochasticity in the data set
    \label{fig:fft_original_example}
\end{figure}

\subsubsection{Error Metrics}
\label{sec:err_metrics}

The statistical error present in the training simulations will propagate to the surrogate data.
Predicting synthetic observables from radiative transfer simulations results in Monte Carlo noise, leading to noisy spectra \citep{bullaPOSSISPredictingSpectra2019, kasenOriginHeavyElements2017}: this represents the floor of the statistical noise.
A surrogate model could recreate the noise present in the training data, but no surrogate model, even a hypothetical model with infinite capacity, will be able to perfectly predict the noise within the unseen test set.
Therefore, the Monte Carlo noise from the radiative transfer simulations leads to a source of error in the final test set predictions.
We would like to know the value of the error that is incurred because of the Monte Carlo noise present in the test data set.
Estimating the true Monte Carlo noise would require simulating the observables for each parameter simulation several times and then computing residuals from the mean spectrum \citep{bullaPolarizationSpectralSynthesis2015}--- a process that is computationally expensive.
To estimate the error, we emulate the mean spectrum by smoothing each spectrum using a Gaussian smoothing method.
We cannot know which value of the Gaussian kernel represents the true mean spectrum best, so we perform Gaussian smoothing for several values of the Gaussian kernel.
We then visually inspect the spectra to check for an approximate best-fit, which occurs at a value of $\sigma=2.5$. 
Fig.~ \ref{fig:fft_original_example} shows an original smoothed spectra and its smoothed counterpart.
In Section \ref{sec:results}, we compare our estimated fractional Monte Carlo noise with the errors of the \texttt{KilonovaNet} predictions.

We evaluate the model performance using three metrics: the spectral error, the bolometric luminosity error, and the band magnitude error. 

The spectral error is the fractional error between the model prediction in a given wavelength bin $y_{\rm pred}(\lambda)$ and the value of the original test data point $y_{\rm test}(\lambda)$:
\begin{equation}
     \epsilon_{s}(\lambda) =  \frac{y_{\rm pred}(\lambda) - y_{\rm test}(\lambda)}{y_{\rm test}(\lambda)}.
\end{equation}

\noindent To connect with photometric survey observations, we quantify the bolometric luminosity error as
\begin{equation}
    \epsilon_b =  \frac{\int y_{\rm pred}(\lambda)d\lambda - \int y_{\rm test}(\lambda)d \lambda}{\int y_{\rm test}(\lambda) d \lambda}.
\end{equation}
%and evaluate it over all the integrated spectra in our test set.

\noindent We also evaluate the performance of the surrogate model by constructing broadband light curves at a distance of 40 Mpc for each of the unique parameter sets in our test data set.
Broadband AB magnitudes from simulated spectra are computed by convolving the flux with the broad-band filters at a chosen distance (40 Mpc in this work).
The distance is chosen to correspond to the distance of GW170817 which allows us to compare with past analyses.
These light curves are computed as AB magnitudes in each band\footnote{We use the LSST filters provided at \url{http://svo2.cab.inta-csic.es/svo/theory/fps3/}}, of which examples are shown in Fig.~ \ref{fig:example_lcs}.
We compute and evaluate the error in the AB magnitude in each band as
\begin{equation}
\Delta m = m_{\rm band, pred} - m_{\rm band, test},
\end{equation}
\noindent where $m_{\rm band,pred}$ is the AB magnitude in a band at a given time step predicted by \texttt{KilonovaNet}, and $m_{\rm band,test}$ is magnitude found from the corresponding spectrum in the test set.

Finally, we will evaluate the performance of the surrogate model by performing a representative inference task and comparing with previously published results. 
We perform parameter estimation using nested sampling with the \texttt{dynesty} \citep{speagleDynestyDynamicNested2020b} sampler on the GW170817 light curve data using our surrogate models.
We compare the differences in the best fit parameters between our fit using the same dataset, which was first collated in \cite{coughlinConstraintsNeutronStar2018} and previously published fits for the same BNS kilonova models but using a different surrogate construction methods. 

%%%%%%%%%%%%%%%%%%%%%%%%%%%%%%%%%%%%%%%%%%%%%%%%%%%%%%%%%%%%%%%%
%%%%%%%%%%%%%%%%%%%%%%%%%%%%%%%%%%%%%%%%%%%%%%%%%%%%%%%%%%%%%%%%
\section{Results}
\label{sec:results}
We present the results of training and evaluating the cVAE on the three simulated datasets. 
For each simulated dataset, we have nine trained models (one for each data-splitting experiment as described in \S~\ref{sec:datapreproc}.
We pass the test data sets --- i.e., the pairs of input ($\mathbf{x}$) and output ($\mathbf{y}$) values that were not used in any aspect of training or model selection --- through the decoders of each model to obtain a predicted spectrum. 
We compare the predictions and the true values of the spectra by computing the values presented in Section \ref{sec:err_metrics}. 

We report a detailed analysis of the surrogate model for the \emph{Dietrich} BNS data set and then report only key values for the \emph{Anand} and \emph{Kasen} data sets; the analysis is the same for all three data sets.
We focus on the \emph{Dietrich} dataset, because we have observational data for a single BNS event, and the \emph{Dietrich} model is a newer BNS kilonova simulation set.

\subsection{Quantifying Uncertainties}

\subsubsection{Spectral Error}
\label{sec:spec_err}

We first compute $\texttt{median}_{\mathbf{x}, \lambda}\epsilon_s(\lambda)$  and $\texttt{mean}_{\mathbf{x}, \lambda}\epsilon_s(\lambda)$ for each of the nine data-split experiments.
We calculate the median and mean over all sets of test input parameters (including the time parameter) and over all wavelengths in the spectra $\mathbf{y}$.
We then compute the mean and standard deviation of the aforementioned means and medians over all nine data-split experiments.

We report the median and mean of the absolute value of $\epsilon(\lambda)_{s}$: $0.067 \pm 0.20$ and $6620 \pm 3180$, where the error is given as the standard deviation over the aforementioned 9 test splits. 
$\texttt{mean}_{\mathbf{x}, \lambda}\epsilon_s(\lambda)$ is heavily skewed by over-predictions when the true value is close to zero for a few outlier values and is therefore not a true representation of the of the typical error.
Although it is not representative, we report the mean so that we may look for skew of the error distribution.
Because the mean is greater than the median, the distribution of errors is skewed high towards over-prediction or slightly brighter kilonovae. 
The corresponding values for the \emph{Anand} and \emph{Kasen} data sets are given in Table \ref{table:all_errs}. 

We then perform the same set of operations on the absolute value of the spectral error:
the mean and variance of this value is $0.285 \pm 0.004$, where the variance is again over all nine experiments. 
We estimate the median absolute fractional Monte Carlo noise using smoothing via a Gaussian kernel with $\sigma=2.5$ and find it to be 0.214.
The size of the median spectral error $\texttt{median}_{\mathbf{x}, \lambda} |\epsilon_s(\lambda)|$ is due to the cVAE surrogate learning the general shape of the spectra but not the Monte Carlo noise.
We leave the interpretation and impact of the size of the error for when we discuss errors in broadband filters.

\begin{figure}
    \centering
    \includegraphics[width=0.45\textwidth]{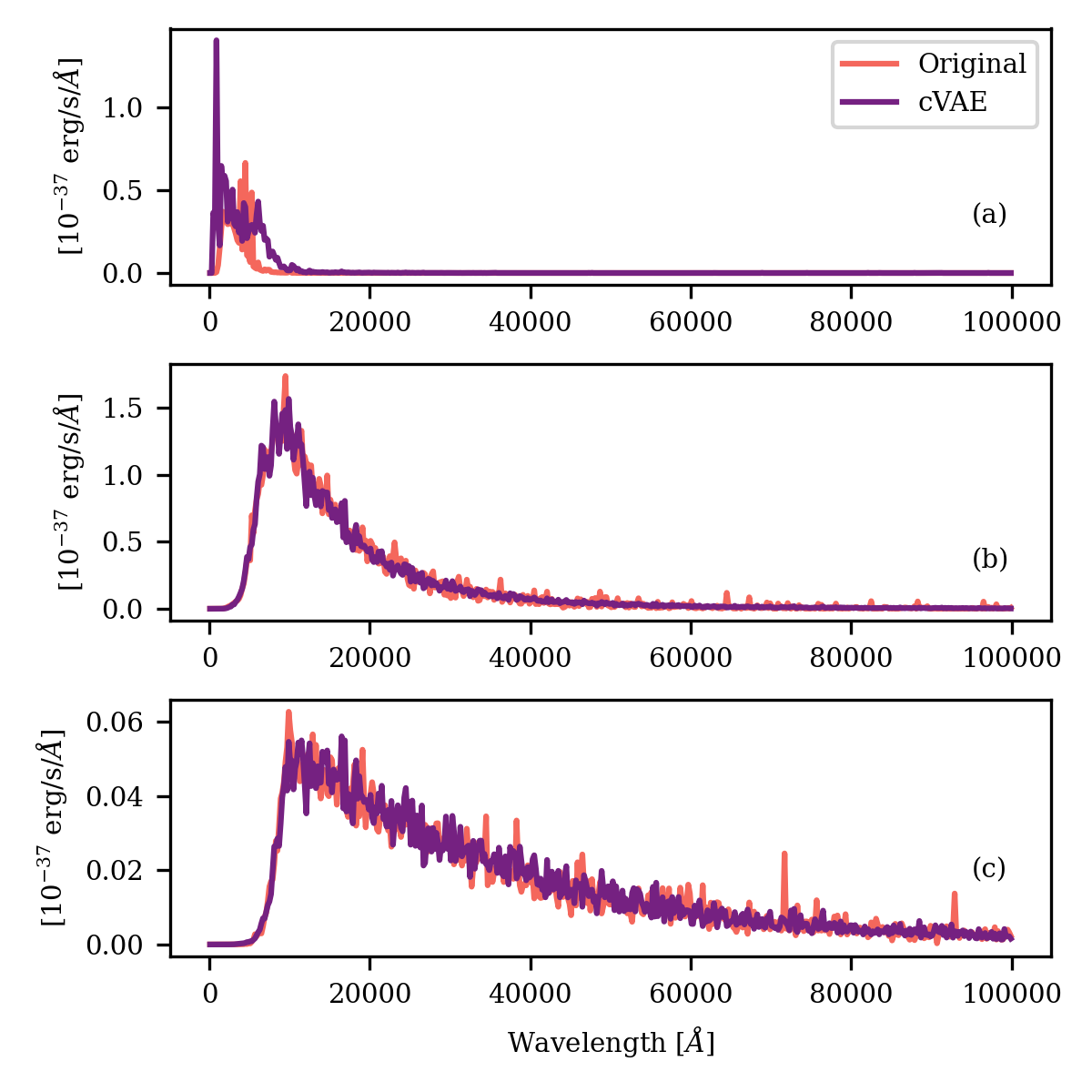}
    \caption{Three original spectra (orange) and corresponding cVAE predictions (blue) for physical parameters  $M_{ej, dyn}/M_{\odot} = 0.02$, $M_{ej, pm}/M_{\odot} = 0.05$, $\Phi = 45.0^{\circ}$, $\cos$ $\Theta_{obs} = 0.8$, and times (a) 0.2 days, (b) 4.2 days, and (c) 14.2 days. The corresponding median spectral errors across the whole spectra (a) 12.91, (b) 0.37,  and (c) 0.20. 
    }
    \label{fig:example_spectra}
\end{figure}

In Fig.~\ref{fig:example_spectra}, we present the spectrum of a BNS kilonova and its corresponding surrogate model predictions for the input parameters $M_{ej, dyn}/M_{\odot} = 0.02$, $M_{ej, pm}/M_{\odot} = 0.05$, $\Phi = 45.0^{\circ}$, and $\cos$ $\Theta_{obs} = 0.8$ at three different time steps (a) 0.2 days, (b) 4.2 days, and (c) 14.2 days. 
The respective $\text{median}_{\lambda}|\epsilon_s|$ for these spectra are 12.91, 0.37, and 0.20.
Fig.~~\ref{fig:example_spectra}a shows significant differences between the original and the surrogate model where the spectrum is non-zero.
In general, we find that all spectra at $t=0.2$ days are poorly predicted by the cVAE.
However, there is general simulation uncertainty at early times due to a lack of atomic data \citep{banerjeeSimulationsEarlyKilonova2020}: due to the lack of detailed simulations of these light curves and the current lack of wide-field, rapid response UV telescopes to detect them, we can safely defer improvement of early-time simulation data to future work.
In contrast, the spectra produced by the cVAE model at $t\gtrsim 1$ day are much more faithful to the original simulations, as evidenced by panels (b) and (c) in Fig.~\ref{fig:example_spectra} and their lower median $|\epsilon_s|$.

\begin{figure*}
    \centering
    \includegraphics[width=0.9\textwidth]{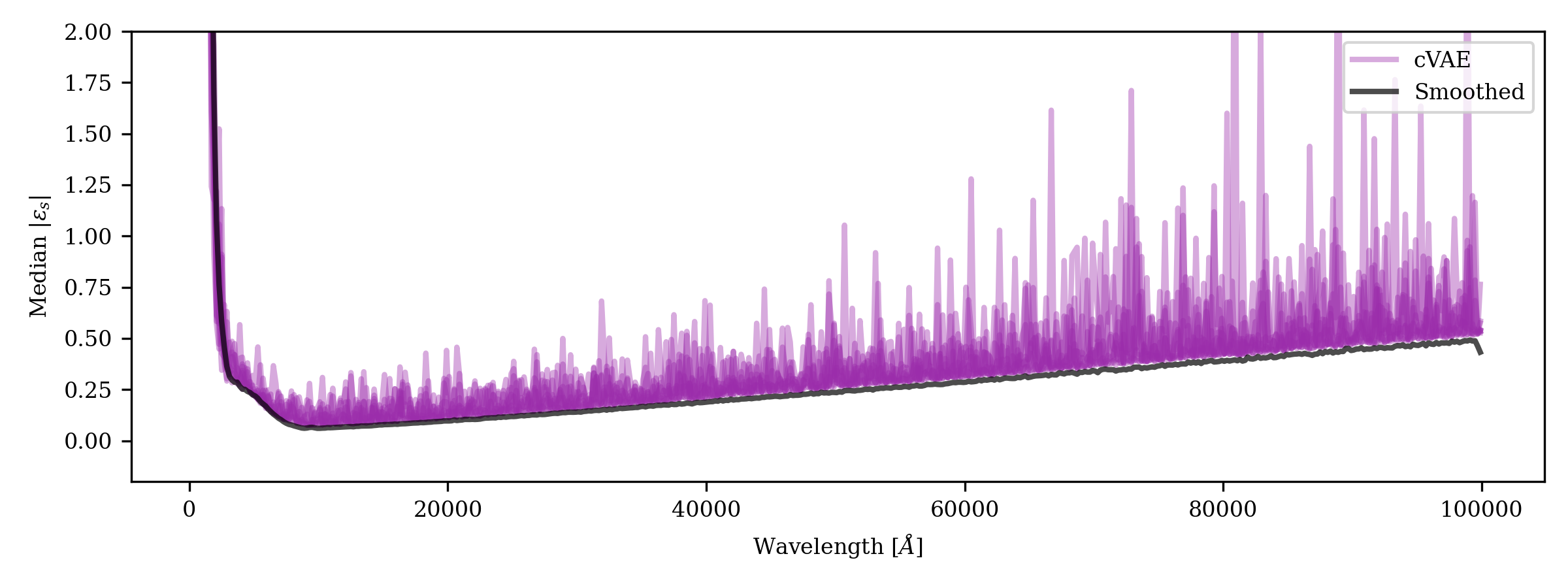}
    \caption{Median of $|\epsilon_s|$ over all the spectra in the \emph{Dietrich} data set for the predictions of the cVAE (purple), along with the absolute spectral error from the Gaussian smoothed spectra for a Gaussian kernel with $\sigma=2.5$ (black).}
    \label{fig:spectral_errors}
\end{figure*}

Fig.~\ref{fig:spectral_errors} shows the median  $\texttt{median}_{\mathbf{x}}|\epsilon_s(\lambda)|$ i.e. the median absolute spectral error over all examples in the test set, including time.
There are nine lines plotted for the medians, corresponding to the nine data split cVAE models.
In addition, we show the estimate of the absolute fractional Monte Carlo noise using smoothing via Gaussian kernel with $\sigma=2.5$ as a function of wavelength. 
The computation of the fractional Monte Carlo noise is the same as the computation of $|\epsilon_s|$ and the values can thus be compared directly.
The Monte Carlo noise estimate traces the bottom of the errors produced by the cVAE;
This could imply that the Monte Carlo noise sets a lower limit for $|\epsilon_s|$ of the cVAE or perhaps any surrogate model.

\begin{table*}
\centering
\caption{Summary of errors for all models. }
\begin{tabular}{lllllll}
\toprule
                                 & $\bm{|\epsilon_s|}$    & \textbf{MC Noise Estimate} & $\bm{\epsilon_{s, mean}}$   & $\bm{\epsilon_{s, med}}$ & $\bm{|\epsilon_b|}$    &                   \\ \toprule
\emph{Dietrich} & $0.285 \pm 0.004$ & 0.214             & $6620 \pm 3180$        & $0.067 \pm 0.20$    & $0.033 \pm 0.005$ &                   \\ 
\emph{Anand}    & $0.292 \pm 0.005$ & 0.241             & $1280000 \pm 95900 $ & $0.114\pm 0.016$    & $0.027 \pm 0.004$ &                   \\ 
\emph{Kasen}    & $0.202 \pm 0.005$ & 0.127             & $0.213 \pm 0.044$      & $0.009 \pm 0.014$   & $0.057 \pm 0.008$ &                   \\ \toprule
                                 & $\bm{|\Delta m|_u}$    & $\bm{|\Delta m|_g}$    & $\bm{|\Delta m|_r}$         & $\bm{|\Delta m|_i}$      & $\bm{|\Delta m|_z}$    & $\bm{|\Delta m|_y}$    \\ \toprule
                                 
\emph{Dietrich} & $0.281 \pm 0.016$ & $0.164 \pm 0.014$ & $0.099 \pm 0.028$      & $0.075 \pm 0.019$   & $0.065 \pm 0.019$ & $0.052 \pm 0.006$ \\ 
\emph{Anand}    & $0.462 \pm 0.016$ & $0.263 \pm 0.016$ & $0.145 \pm 0.009$      & $0.101 \pm 0.013$   & $0.088 \pm 0.014$ & $0.081 \pm 0.009$ \\ 
\emph{Kasen}    & $0.176 \pm 0.007$ & $0.136 \pm 0.014$ & $0.094 \pm 0.011$      & $0.090 \pm 0.015$   & $0.083 \pm 0.010$ & $0.092 \pm 0.010$ \\
\bottomrule
\end{tabular}
\label{table:all_errs}
\end{table*}

\subsubsection{Bolometric Luminosity Error}
Next, we discuss the errors in the bolometric light curves.
For \emph{Dietrich}, $\texttt{mean}_{\mathbf{x}}|\epsilon_b| = 0.033 \pm 0.005$, where the error is the standard deviation across all nine data split experiments.

\begin{figure}
    \centering
    \includegraphics[width=0.45\textwidth]{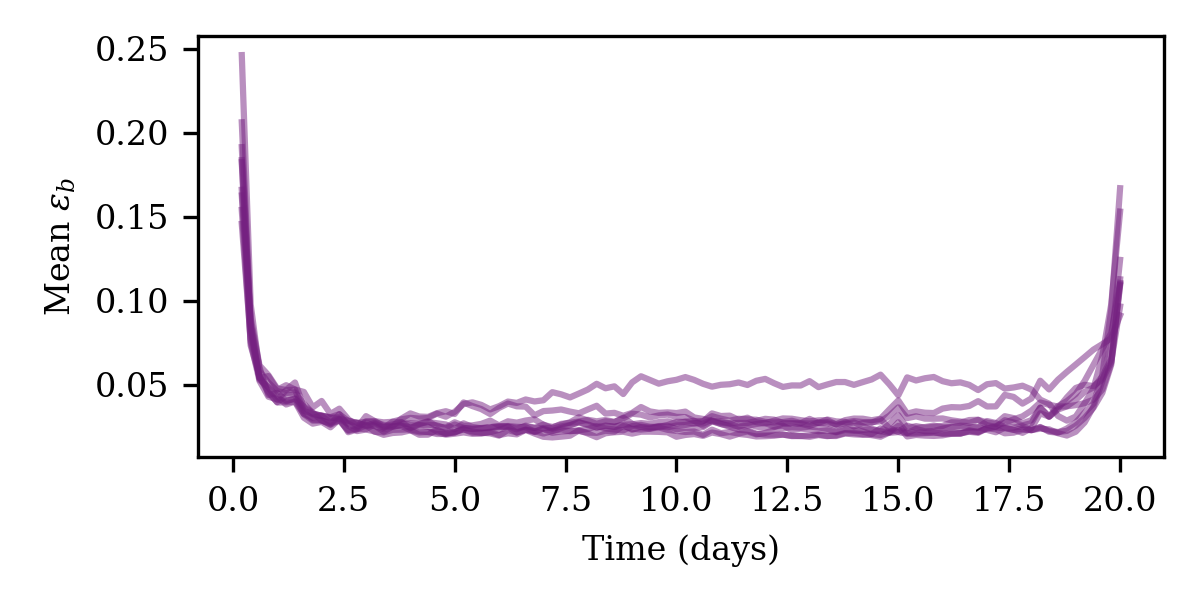}
    \caption{Mean of $|\epsilon_b|$ across all nine data-split models of the \emph{Dietrich} dataset. 
    % Mean values across the test BNS dataset for the bolometric luminosity error for all nine models. 
    The two times at which the model is least accurate is at $t=0.2$ and $20$ days.}
    \label{fig:bolometric_v_time}
\end{figure}

An investigation into the outliers in the \emph{Dietrich} dataset shows that the spectra with the largest bolometric luminosity error occur at times $t=0.2$ and $t=20$.
The spectra with the highest bolometric luminosity error (of 1.56) is shown in Fig.~\ref{fig:example_spectra}a, which also corresponds to $t=0.2$.
The spectra at $t=20$ are also often over-predicted, leading to $\epsilon_b$ values of approximately 0.7.
The outliers at $t=0.2$ and $t=20$ are the most prominent features across all nine data-split models in the plot of mean bolometric luminosity error $\text{mean}_{M_{ej, dyn}, M_{ej, pm}, \Phi, \cos \theta_{obs}} | \epsilon_b|$ as a function time, which is shown in Fig.~ \ref{fig:bolometric_v_time}.
The correlations with time are different for the other two datasets.
The light curves of the \emph{Kasen} set have a tendency to quickly fall to zero after $\sim$14 days: the errors increase rapidly at about that time.
The \emph{Anand} dataset also exhibits a large error peak at $t=0.2$, but since BHNS kilonovae dim more slowly, the errors remain consistently low even at $t=20$ days.

\begin{figure*}
    \centering
    \includegraphics[width=0.9\textwidth]{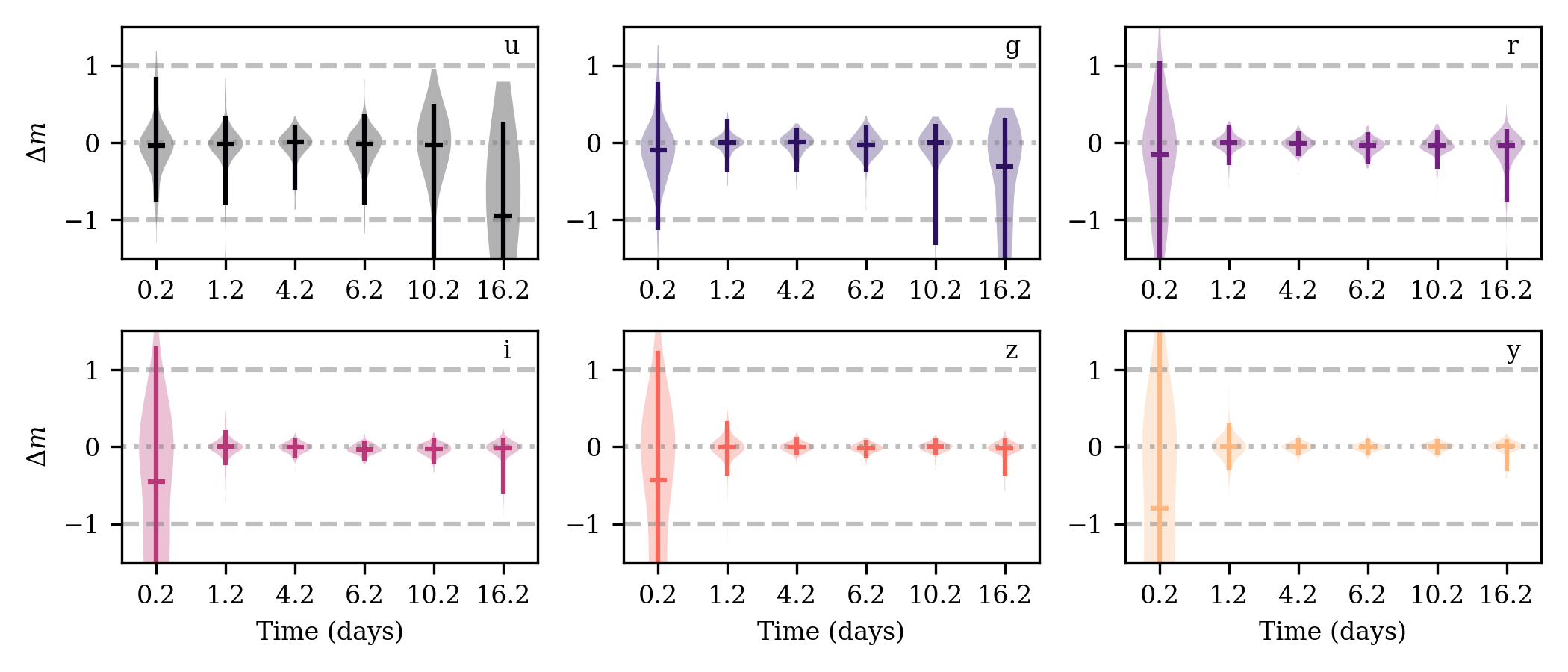}
    \caption{Distributions of magnitude errors $\Delta m$ between the predictions and the test dataset for five time steps in the \emph{Dietrich} simulation set. 
    The horizontal lines for each violin represents the median of the distribution, and the endpoints of the vertical lines represent the 95\% confidence interval.}
    \label{fig:delta_mag_violins}
\end{figure*}

\subsubsection{Broadband Magnitude Error}
Finally, we evaluate the performance of the surrogate model in broadband magnitudes, $ugrizy$. 
We construct light curves for each set of input parameters in the test dataset and compute the mean magnitude differences. 
We report the mean of the magnitude differences $\texttt{mean}_{\mathbf{x}}\Delta m$, where a mean is taken across all sets of input parameters in the test set, where time is also treated as an input parameter.
We report the values for \emph{Kasen} and \emph{Anand} in Table \ref{table:all_errs}.

For the \emph{Dietrich} test set, we additionally show the distribution of $\texttt{mean}_{M_{ej, dyn}, M_{ej, pm}, \Phi, \cos \theta_{obs}} \Delta m$ for selected time steps of the data set.
Fig.~\ref{fig:delta_mag_violins} shows these distributions of $\Delta m$ for all six bands.
In general, for the time steps other than $t=0.2$, the width of the distributions of $\Delta m$ is correlated with brightness. 
Overall, most of the predictions are within 0.1 mag of the test data set.
This is well within the commonly assumed 1 mag uncertainty in other KN light curve modeling fitting  \citep{coughlinConstraintsNeutronStar2018}.
The least accurate predictions occur for the \emph{Anand} dataset in $u$-band, with a median difference of 0.451 mag.
Since the \emph{Anand} dataset is of a BHNS model, the model does not contain a lanthanide-free (``bluer") component and would thus be practically undetectable in the blue bands, leading to high errors in the surrogate model.

\begin{figure*}
    \centering
    \includegraphics[width=0.9\textwidth]{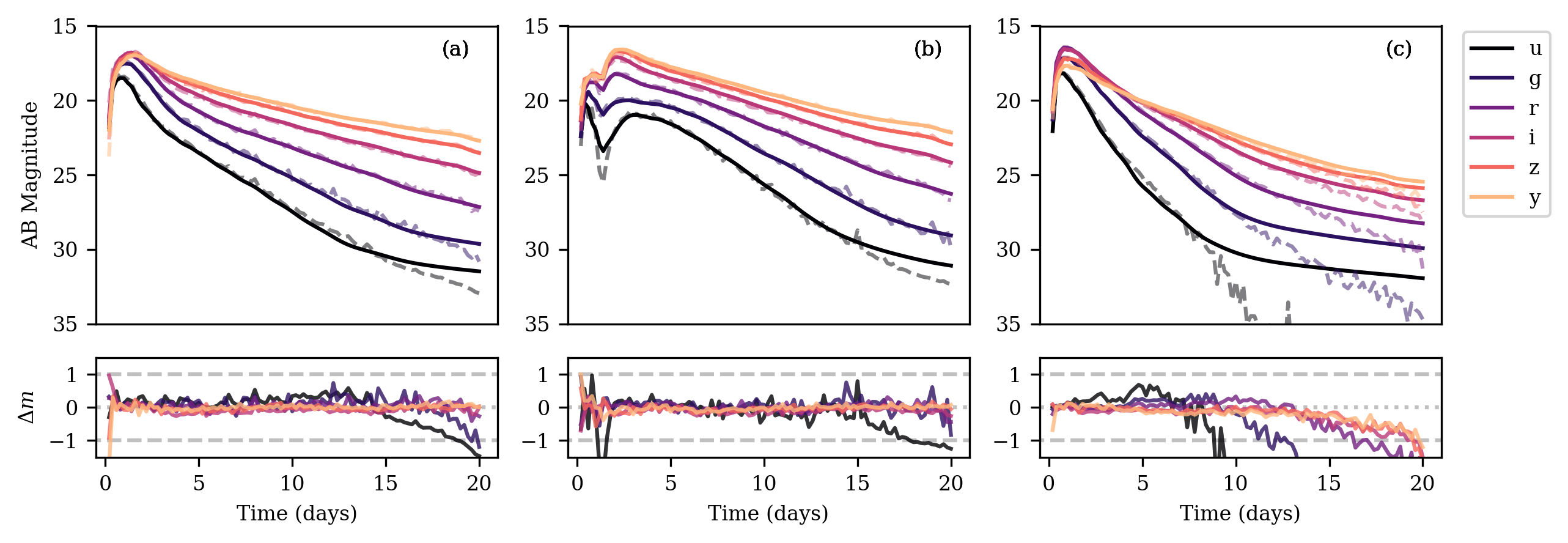}
    \caption{Three example light curves from the \emph{Dietrich} BNS datasets for the original dataset (dashed lines) and the cVAE predictions (solid lines) for the six $ugrizy$ bands (colors black through yellow), as well as the residuals between original and prediction for each lightcurve and band. 
    The three parameter sets are 
    (a) $M_{ej, dyn}/M_{\odot}=0.01$, $M_{ej, pm}/M_{\odot}=0.09$, $\Phi$= 30.0, $\cos \Theta_{obs}$=0.3, 
    (b) $M_{ej, dyn}/M_{\odot}=0.001$, $M_{ej, pm}/M_{\odot}=0.11$, $\Phi=75.0$, $\cos\Theta_{obs}=0.0$,
    (c)$M_{ej, dyn}/M_{\odot}=0.02$, $M_{ej, pm}/M_{\odot}=0.01$, $\Phi$: 0.0, $\cos\Theta_{obs}$=0.3. 
    The lower panels indicate the  differences between the predicted light curve and the original as a function of time, for each light curve.}
    \label{fig:example_lcs}
\end{figure*}

We present three examples for light curves produced by the cVAE, along with the original light curves and absolute differences between the prediction and original in Fig.~ \ref{fig:example_lcs}.
The three randomly chosen examples are representative of the remainder of the test set in that they show that the cVAE predictions in bands are smooth and that the predictions become unreliable when the light curve is too faint.
Fig.~ \ref{fig:example_lcs}c shows an example for a model free of lanthanide rich ejecta, since the half opening angle $\Phi$ of the equatorial component is zero.
This particular example peaks quickly and dims quickly, and below mag$\sim27$, the cVAE fails to predict the original data.
However, the kilonova at a distance of 40 Mpc would not be detectable at such a magnitude; for example, the design single-visit limiting magnitude for the LSST $u$-band is 23.9 \citep{ivezicLSSTScienceDrivers2019}.
Fig.~ \ref{fig:example_lcs}c shows the limitations of the model, especially at the edges of the parameter space, but we emphasize that that the errors primarily occur when the kilonova has dimmed beyond detectability.

\subsection{Fitting AT2017gfo}
We test the performance of the surrogate model in the scientific context that it will be used: we use the \emph{Dietrich}-based surrogate model to perform kilonova parameter estimation on the object AT2017gfo (counterpart to the event GW170817) with the data collated in \cite{coughlinConstraintsNeutronStar2018} from the sources of \citet{andreoniFollowGW170817Its2017, arcaviOpticalEmissionKilonova2017,chornockElectromagneticCounterpartBinary2017, cowperthwaiteElectromagneticCounterpartBinary2017,droutLightCurvesNeutron2017,evansSwiftNuSTARObservations2017,kasliwalIlluminatingGravitationalWaves2017,tanvirEmergenceLanthaniderichKilonova2017,pianSpectroscopicIdentificationRprocess2017,trojaXrayCounterpartGravitationalwave2017,smarttKilonovaElectromagneticCounterpart2017,utsumiJGEMObservationsElectromagnetic2017}.
We use flat priors that extend over the published data range and therefore the range of supported parameter by the surrogate model $\log_{10}(0.001) \leq \log_{10}(M_{ej, dyn} / M_{\odot}) \leq \log_{10}(0.02)$, $\log_{10}(0.01) \leq \log_{10}(M_{ej, pm} / M_{\odot}) \leq \log_{10}(0.13)$, $0 \leq \cos (\Phi) \leq 1$, and $0 \leq \cos (\theta) \leq 1$.

\begin{figure}
    \centering
    \includegraphics[width=0.35\textwidth]{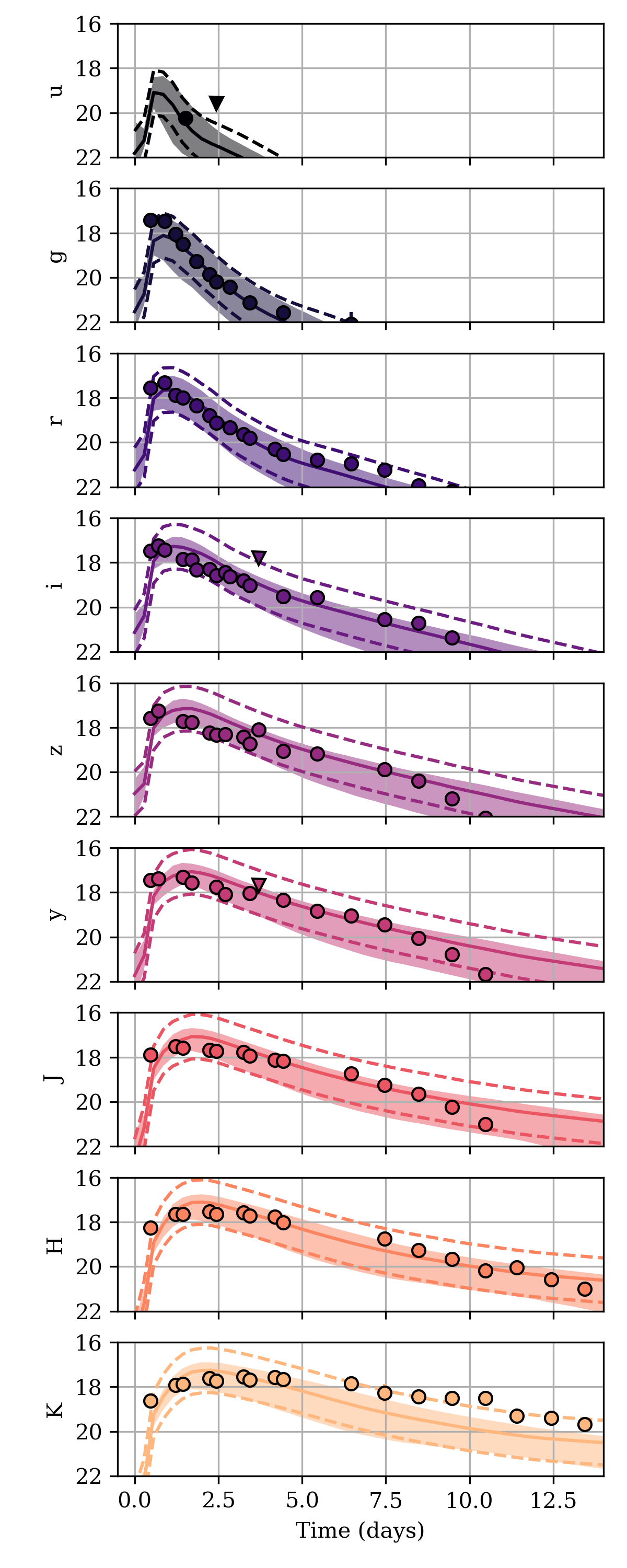}
    \caption{
    Light curves for AT2017gfo. 
    Observed values (points and triangles, where the latter are lower bound observations) and the prediction based inferred parameters using the \emph{Dietrich}-based cVAE surrogate model (solid lines). 
    The shaded bands represent the 90 \% confidence interval of light curves constructed from the posterior samples. 
    The dashed lines represent the 1 mag tolerance typically used to represent modelling error of kilonova light curves. }
    \label{fig:fit_lightcurve}
\end{figure}

\begin{figure*}
    \centering
    \includegraphics[width=0.9\textwidth]{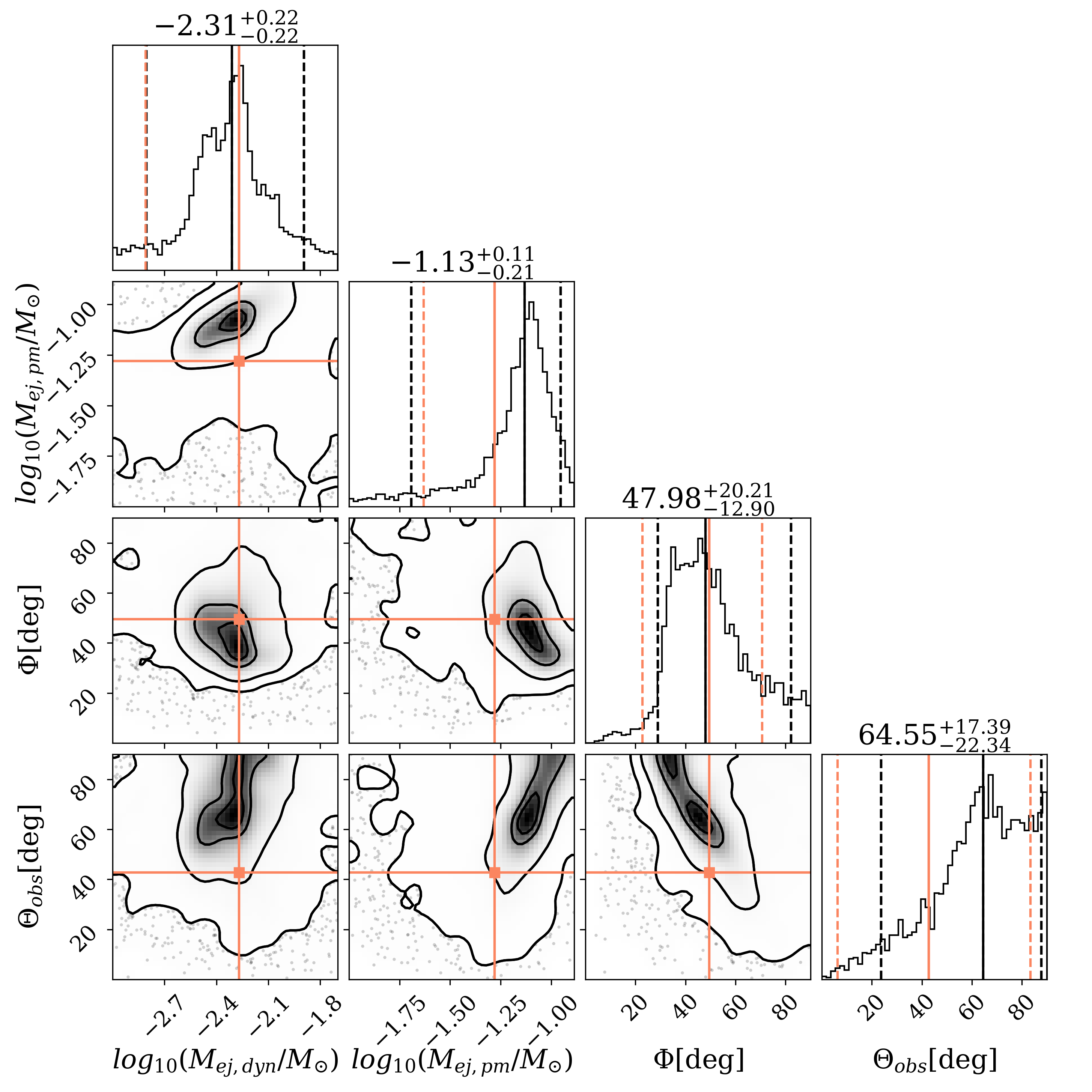}
    \caption{Inferred posteriors of model parameters  $M_{ej, dyn}$, $M_{ej, pm}$, $\Phi$, and $\cos \theta_{obs}$ from the observations of AT2017gfo (as collated in Coughlin et al. (2018)) at 10\%, 32\%, 68\%, and 95\% confidence intervals. The median values and 90\% confidence intervals are shown as vertical solid and dashed lines and above each column, where the orange lines indicate the 90\% confidence interval from   \protect\cite{dietrichMultimessengerConstraintsNeutronstar2020} using the same dataset and kilonova model but a different surrogate construction and prior and the black lines indicate our 90\% confidence interval.}
    \label{fig:fit_corner}
\end{figure*}

We use the \texttt{dynesty} sampler and the log-likelihood (up to a constant) of
\begin{equation}
    \log\mathcal{L} = -\frac{1}{2} \sum_{i=1}^N \frac{(m_{i, pred} - m_{i, obs})^2}{\sigma_{i, obs}^2 + \sigma_{ sys}^2},
    \label{eq:lklhd}
\end{equation}
\noindent where $i$ indexes observations, $m_{i,pred}$ is the magnitude of each proposed sample, $m_{i,obs}$ is the magnitude of each observation, and $\sigma_{i,obs}$ is the uncertainty of each observation. 
The likelihood is Gaussian with an additional systematic uncertainty of $\sigma_{sys}=1$ mag added to account for the modelling uncertainty, as used in \cite{coughlinConstraintsNeutronStar2018}. 
Fig.~\ref{fig:fit_lightcurve} shows the light curves derived from the fit, and Fig.~\ref{fig:fit_corner} shows the parameter posteriors, as well as the best-fit median values on the same data set obtained by \cite{dietrichMultimessengerConstraintsNeutronstar2020}. 

There are some key differences to note for this comparison.
The fit performed by \cite{dietrichMultimessengerConstraintsNeutronstar2020} on the light curve data uses as a prior gravitational wave and pulsar observations. 
In addition, they used different allowed ranges of the kilonova model parameters.
With all these differences in mind, the agreement between our parameter recovery and the fit presented in their analysis shows remarkable consistency, with each of the medians of their recovered parameters lying within $1 \sigma$ of our median recovered parameters, as shown in Fig.~\ref{fig:fit_corner}. 
The parameter inference performed with \texttt{dynesty} for this data set required 3.5 minutes on an Apple M1 Pro chip.

In the Appendix, we include a more directly comparable fit using the \emph{Kasen} model as well. 
An additional test would involve generating light curves from injected parameters and seeing whether the same parameters are recovered through a sampling-based fit.
This has been performed in \cite{lukosiute_2021} using the surrogate constructions described in this work for the \emph{Dietrich} and \emph{Anand} data sets with success; see \cite{lukosiute_2021} for details. 

\section{Conclusion}
\label{sec:conclusion}
In this work, we presented an implementation of the conditional variational autoencoder (cVAE) as a method for surrogate model construction of kilonova spectra. 
We discussed the method's potential to produce complex, high-dimensional data, including the theoretical and practical limitations, especially with respect to uncertainty quantification.
We applied the method to three commonly used kilonova spectra model datasets ---  those published in \cite{kasenOriginHeavyElements2017}, \cite{dietrichMultimessengerConstraintsNeutronstar2020}, and \cite{anandOpticalFollowupNeutron2021}. 
Because the model is unable to learn its own uncertainties, we delineated potential sources of error and developed a suite of metrics tailored to these specific data sets and their scientific uses. 

We evaluated the surrogate model's performance in the context of directly predicting spectra, as well as on downstream products like the bolometric luminosities and {\it ugrizy} broadband magnitudes. 
As a final test of the model, we applied it in a typical scientific use case: measuring the physical parameters of the kilonova associated with GW170817.
While the cVAE method is limited by variance shrinkage, it can still quickly produce spectra with high fidelity to the original training simulations. 
Kilonova models that are published in the future can have a surrogate pre-trained and be used for fast inference for any set of photometric bands.
 
We leave for future work studying how to address the variance shrinkage problem so that the variance of the surrogate can be learned and the uncertainty incorporated into Bayesian analyses.
Some potential avenues include using the CombVAE \citep{detlefsenReliableTrainingEstimation2019} or the QR-VAE \cite{akramiAddressingVarianceShrinkage2020}, or other machine learning methods that allows for accurate density estimation, such as normalizing flows \cite{rezendeVariationalInferenceNormalizing2016, dinhNICENonlinearIndependent2015},.
We leave also for future work how to incorporate the errors incurred due to surrogate modelling into scientific analyses. 
Nevertheless, we have shown that the cVAE can already serve as a fast and accurate surrogate model for kilonova spectra and have evaluated uncertainty without modeling variance directly.
We release the final trained models for the three datasets and user-friendly code to produce spectra at \url{https://github.com/klukosiute/kilonovanet}.

\section{Acknowledgments}
We acknowledge the contributions of the authors in detail.

K.~Lukosiute led all computations and writing.
G.~Raaijmakers consulted on analysis of the observational data and co-wrote.
Z.~Doctor consulted on the statistical analysis and co-wrote.
M.~Soares-Santos posed the problem and edited the manuscript.
B.~Nord helped develop the initial computational approach, co-designed the statistical analysis, and co-led writing of the paper.

We also thank Samaya Nissanke, Bas Dorsman, and James Annis for useful discussions and comments. 

Z.D.~is supported by the CIERA Board of Visitors Research Professorship.
G.R. and K.L. are grateful for financial support from the Nederlandse Organisatie voor Wetenschappelijk Onderzoek (NWO) through the Projectruimte and VIDI grants (Nissanke).
We acknowledge the Deep Skies Lab as a community of multi-domain experts and collaborators who’ve facilitated an environment of open discussion, idea-generation, and collaboration. This community was important for the development of this project.
We thank SURFsara (www.surfsara.nl) for the support in using the Lisa Compute Cluster.

Work supported by the Fermi National Accelerator Laboratory, managed and operated by Fermi Research Alliance, LLC under Contract No. DE-AC02-07CH11359 with the U.S. Department of Energy. The U.S. Government retains and the publisher, by accepting the article for publication, acknowledges that the U.S. Government retains a non-exclusive, paid-up, irrevocable, world-wide license to publish or reproduce the published form of this manuscript, or allow others to do so, for U.S. Government purposes.

\bibliographystyle{mnras.bst}
\bibliography{bibliography.bib}

\section*{Appendix}
\begin{figure}
    \centering
    \includegraphics[width=0.45\textwidth]{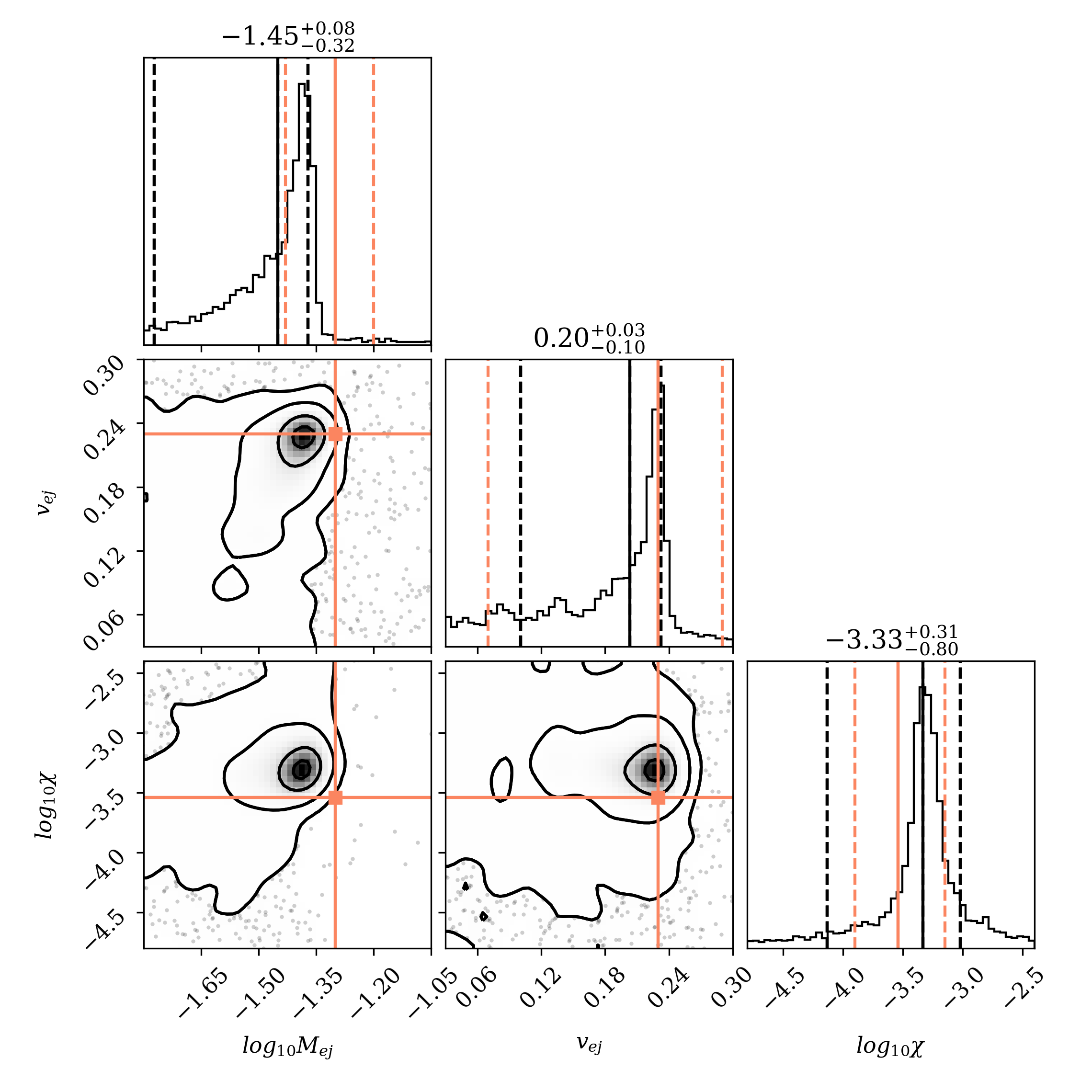}
    \caption{
    Corner plot for the inferred posterior distribution of the \emph{Kasen} model parameters mass of the ejecta $M_{ej}$, velocity of the ejecta $v_{ej}$, and lanthanide fraction $\chi$ from the observations of AT2017gfo as collected in \citet{coughlinConstraintsNeutronStar2018} at 10\%, 32 \%, 68 \%, and 95\% confidence. 
    The dashed lines in the 1D distributions, as well as the values given in the titles, represent 68 \% confidence interval, with the median lying between the two dashed lines representing the median value. 
    The orange solid lines indicate the median values from the results of the fit performed in \citet{coughlinConstraintsNeutronStar2018} using the same dataset and kilonova model but a different surrogate construction and prior. 
    The orange dashed lines indicate the ranges presented in \citet{coughlinConstraintsNeutronStar2018}. }
    \label{fig:fit_corner_kasen}
\end{figure}

\begin{figure}
    \centering
    \includegraphics[width=0.35\textwidth]{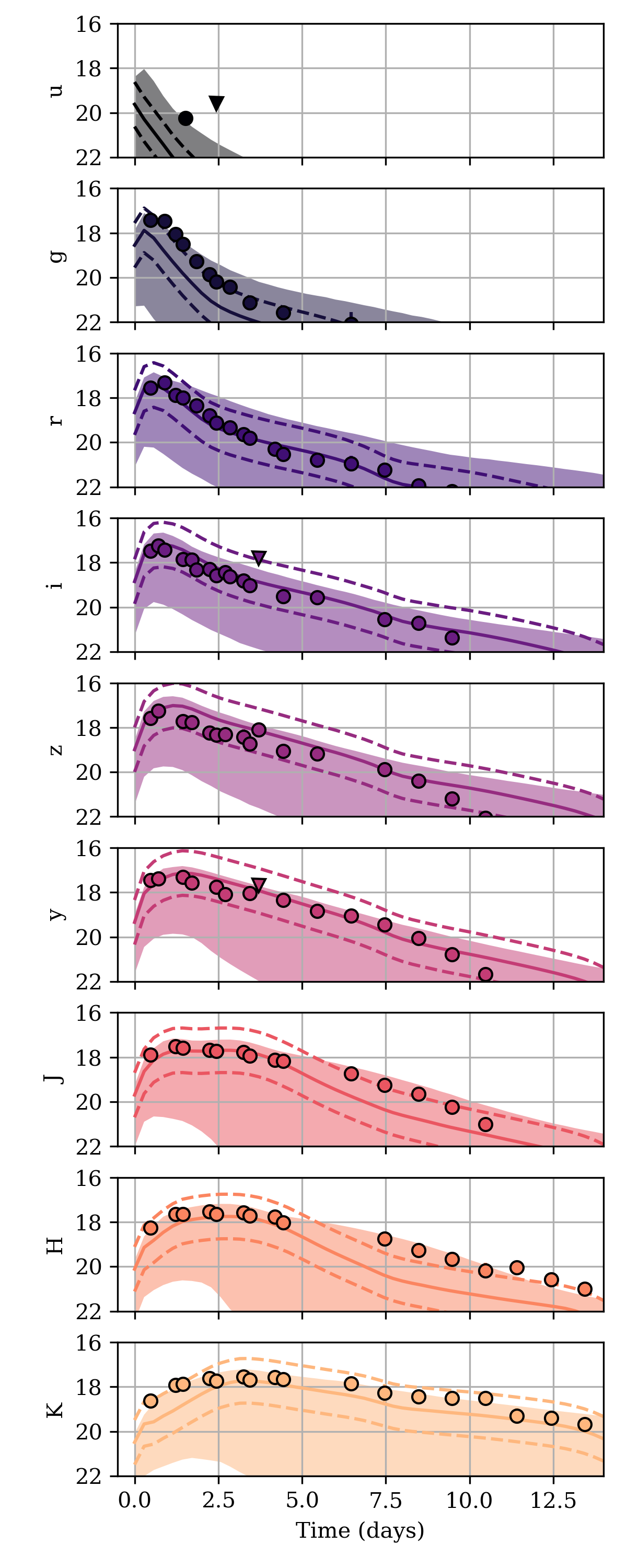}
    \caption{Comparison of the observed lightcurve for AT2017gfo (points) and the prediction from inferred parameters using the cVAE surrogate model (solid lines) using the \emph{Kasen} model. The shaded bands represent the 90 \% confidence interval of lightcurves constructed from the posterior samples. The dashed lines represent the 1 magnitude tolerance often used to represent modelling error of kilonova light curves.}
    \label{fig:fit_lightcurve_kasen}
\end{figure}

In addition to performing parameter estimation for AT2017gfo for the \emph{Dietrich} model, we also perform parameter estimation for the single component \emph{Kasen} model.
Most literature agrees that more than one component is present in kilonova outflows \citep{raaijmakersChallengesAheadMultimessenger2021, dietrichModelingDynamicalEjecta2017, kasenOPACITIESSPECTRATHErPROCESS2013, coughlinConstraintsNeutronStar2018}.
We aim to compare the performance of our surrogate model directly to another published result.
Therefore, we present this particular parameter estimation to compare with the parameter inference presented in \cite{coughlinConstraintsNeutronStar2018}.
We use the same likelihood (Equation \ref{eq:lklhd}), sampling software (\texttt{dynesty}), and observational dataset for the lightcurve of AT2017gfo as we used for the parameter inference for the \emph{Dietrich} model  \citep{speagleDynestyDynamicNested2020b}.
We use flat priors: $\log_{10}(0.001) \leq \log_{10}(M_{ej, dyn} / M_{\odot}) \leq \log_{10}(0.1)$, $0.03c \leq v_{ej} \leq 0.3c$, and $-9 \leq  \log_{10}(\chi) \leq -1$.
The results for our fit are shown in Fig.~ \ref{fig:fit_corner_kasen} and \ref{fig:fit_lightcurve_kasen}.
We find relatively good agreement between our inferred parameters and those presented in \cite{coughlinConstraintsNeutronStar2018}: all of their parameters fall within $1\sigma$ of our parameters.
Some of the disagreement in fit is likely explained by the slight differences in the data that our surrogate was trained on and the prior ranges.
We use both the systematic parameter survey and the narrower survey and combine them for our training set to construct the surrogate for the \emph{Kasen} model, so our surrogate model uses more data than the surrogate model used in \cite{coughlinConstraintsNeutronStar2018}.
Additionally, \cite{coughlinConstraintsNeutronStar2018} use a wider prior that extends outside the ranges of the kilonova models published in \cite{kasenOriginHeavyElements2017}.

\end{document}